\documentclass{article}%
\usepackage{amsmath}
\usepackage{amsfonts}
\usepackage{amssymb}
\usepackage{graphicx}%
\providecommand{\U}[1]{\protect\rule{.1in}{.1in}}
\begin{document}

\begin{center}
\textbf{BRANS-DICKE WORMHOLE REVISITED-II}

\bigskip

Ramil Izmailov,$^{1,a}$ Amrita Bhattacharya$^{2,b}$ and Kamal K.
Nandi$^{1,2,c}$

$\bigskip$

$^{1}$Joint Research Laboratory, Bashkir State Pedagogical University, Ufa
450000, Russia

\bigskip$^{2}$Department of Mathematics, University of North Bengal, Siliguri
734 013, India

\end{center}

\bigskip

\bigskip


\begin{center}
$^{a}$E-mail: izmailov.ramil@gmail.com

$^{b}$E-mail: amrita\_852003@yahoo.co.in

$^{c}$E-mail: kamalnandi1952@yahoo.co.in
\end{center}

\bigskip

\begin{center}
---------------------------------------------------------------------------

\textbf{Abstract}
\end{center}

In a recent paper, a wormhole range in the Jordan frame, $-3/2<\omega<-4/3$
for the vacuum Brans-Dicke Class I solution was derived. On general grounds
and under certain conditions, it is shown in a theorem that static wormhole
solutions in the scalar-tensor theory are not possible. We agree with the
conclusion within its framework but report that a singularity-free wormhole
\textit{can} be obtained from Class I solution by performing certain
operations on it, a fact possibly not yet widely known. The transformed
solution is regular everywhere, produces a wormhole with two asymptotically
flat regions for a revised new range $-2<\omega<-3/2$ , together with a
wormhole analogue (of Horowitz-Ross naked black hole) that we discovered
earlier. This new range lies in the ghost regime in the Einstein frame
consistent with the theorem. We further conclude that there is a fixed point
at $\omega=-3/2$, the values $\omega>-3/2$ correspond to singular wormholes,
while values $\omega<-3/2$ correspond to singularity-free wormholes.

\begin{center}
---------------------------------------------------------------------------
\end{center}

\textbf{I. Introduction}

Lorentzian wormholes as possible astrophysical objects has been under active
investigation for quite some time now. In particular, the possibility of
occurrence of such objects in the Brans-Dicke theory is quite intriguing since
it is a natural theory that emerged as a Machian alternative to Einstein's
theory of general relativity. To our knowledge, a\ theoretical search for
static wormholes in the vacuum Brans-Dicke theory has been initiated by Agnese
and La Camera [1] who have shown that the Brans-Dicke scalar $\varphi$ can
play the role of exotic matter provided the coupling parameter $\omega<-2$,
followed by the work of Visser and Hochberg [2], and by the works in other
classes of Brans solutions [3] in the Jordan Frame(JF or \textbf{M}%
$_{\text{J}}$) as well as in the conformally rescaled Einstein Frame (EF or
\textbf{M}$_{\text{E}}$) [4]. There exist hundreds of articles on wormholes
today, but we only mention some works on \ Brans-Dicke wormholes [5-15]. A
particularly interesting recent result is that the Horowitz-Ross naked black
hole has a wormhole analogue in the Brans Class I solution [13]. Considering
the importance of Brans-Dicke theory in the interpretation of various
astrophysical phenomena, it is important that certain questions raised in the
literature relating to static spherically symmetric wormhole solutions in the
vacuum theory be clarified. The purpose of the present paper is to do that.

For static spherically symmetric Class I solution in the $\omega=$ constant
vacuum Brans-Dicke theory, we proposed a wormhole range $-3/2<\omega<-4/3$ in
the JF and developed a wormhole analogy to Horowitz-Ross naked black holes
[13] for $\omega<-2$, the range obtained previously by Agnese and La Camera
[1]. Against this proposal, we identified two criticisms: (A) The range
$-3/2<\omega<-4/3$ \ lies in the so called "no-ghost" regime in
EF\footnote{Private correspondence based on [14]. By ghost matter, we mean
here matter that violates energy conditions, at least the Null Energy
Condition (NEC). The values of $\omega$ in the scalar field redefinition (4)
determine the sign of the resulting kinetic term in EF action. When $\alpha
<0$, $\omega<-3/2$, the sign is negative and $\omega$ is said to be in the
ghost regime. See Sec.II.} and (B) The occurrence of naked singularity in the
Class I solution spoils the spacetime [15, p.5]. On general grounds, it is
proved in a theorem [14, pp.3,6] that, under certain conditions, there is no
wormhole connecting two spatial infinities in the scalar-tensor theory. In
[14,15], it is of course implied, and we accept, that a true wormhole is meant
to be a non-singular one with two asymptotically flat regions connected by a
throat. A singular wormhole is "diseased" [16]; in fact it is long recognized
that the Class I solution is plagued by such curvature
singularity\footnote{Professor Starobinsky, in another correspondence, has
pointed out an additional problem: The existence of a throat for any value of
$\omega>-3/2$ does not prevent the formation of a curvature singularity behind
it. The difference between the $\omega>-4/3$ and $\omega<-4/3$ cases is only
that the singularity lies at finite or infinite proper radial distance from
the throat, but there is no asymptotic flatness (no "another universe") behind
the throat in both cases. }. In this work, we are concerned only with the
special case of Class I solution in the $\omega=$ constant vacuum Brans-Dicke theory.

We agree with the theorem within the framework it is proved. When applied to
the special case under consideration, its message may be summarized in the
paradigm "no ghost matter $\Rightarrow$ no wormhole" in either frame connected
by conformal mapping [14, p.3]. However, conformal mapping does \textit{not}
always guarantee that ghost-free matter in EF corresponds to ghost-free matter
in JF and vice versa. Our earlier work in fact showed that the ghost-free
regime $\omega>-3/2$ in EF can still lead to ghost-matter in JF [13] together
with the existence of a real throat and twice asymptotically flat regions
(inversion invariant under $r\rightarrow B^{2}/r$). In spite of these
features, we understand that Class I solution does not qualify as a true
wormhole. One may say that $r=0$ flat spatial infinity is divided from the one
at $r=\infty$ by a curvature singularity at $r=B$. So, the inversion
transformation in fact relates two separate singular space-times, not two
regions of one \textit{connected }space-time required of a true wormhole.
Having said this, we argue that the theories in JF and EF are two
fundamentally distinct theories and hence it seems more logical to draw
conclusions about wormhole $\omega-$regime in JF from the geometry in JF itself.

The curvature singularity in Class I solution (in its usual form) is well
known. We show here that by some operations the maladies in the Class I
solution can be redressed leading to an asymptotically flat, singularity-free
wormhole in the new range $-2<\omega<-3/2$. We also show that a gauge
non-uniqueness of solutions allow us to shift the values $\omega$ on either
side of the divide $\omega=-3/2$ in the vacuum JF Brans-Dicke theory. We take
$16\pi G=c=1$ and a signature convention $(-,+,+,+)$.

\textbf{II. Two distinct theories}

For the ease of argument, we restate the actions of two distinct theories.
First is the action of the vacuum Brans-Dicke theory in the JF ($g_{\mu\nu
},\varphi$):
\begin{equation}
S=\int d^{4}x(-g)^{\frac{1}{2}}\left[  \varphi\mathbf{R}+\omega\varphi
^{-1}g^{\mu\nu}\varphi_{,\mu}\varphi_{,\nu}\right]  .
\end{equation}
The field equations are%
\[
\square^{2}\varphi=0,
\]

\begin{equation}
\mathbf{R}_{\mu\nu}-\frac{1}{2}g_{\mu\nu}\mathbf{R}=-\frac{\omega}{\varphi
^{2}}\left[  \varphi_{,\mu}\varphi_{,\nu}-\frac{1}{2}g_{\mu\nu}\varphi
_{,\sigma}\varphi^{,\sigma}\right]  -\frac{1}{\varphi}\left[  \varphi
_{,\mu;\nu}-g_{\mu\nu}\square^{2}\varphi\right]  ,
\end{equation}
where $\square^{2}\equiv(\varphi^{;\rho})_{;\rho}$ and $\omega=$ constant is a
dimensionless coupling parameter.

Consider the conformal transformation%
\begin{equation}
\widetilde{g}_{\mu\nu}=\varphi g_{\mu\nu},\text{ \ \ }%
\end{equation}
and a redefinition of the Brans-Dicke scalar $\varphi\rightarrow\phi$ by
\begin{equation}
d\phi=\left(  \frac{\omega+3/2}{\alpha}\right)  ^{1/2}\frac{d\varphi}{\varphi
},
\end{equation}
in which we have intentionally introduced an arbitrary constant parameter
$\alpha$. These transformations are known for long in the literature as Dicke
transformations. The resulting action in the EF ($\widetilde{g}_{\mu\nu},\phi
$) is%
\begin{equation}
S=\int d^{4}x(-\widetilde{g})^{1/2}\left[  \widetilde{\mathbf{R}}%
+\alpha\widetilde{g}^{\mu\nu}\phi_{,\mu}\phi_{,\nu}\right]  .
\end{equation}
This is the Hilbert-Einstein action of Einstein's general relativity with a
source kinetic term $\alpha\widetilde{g}^{\mu\nu}\phi_{,\mu}\phi_{,\nu}$ that
leads to Einstein minimally coupled field equations. As one sees, $\omega$ has
disappeared from the action. To ensure ghost matter needed for wormholes, one
has to have a \textit{negative} kinetic term. The common prescription is to
assume $\alpha<0$ and a real $\phi$ in EF, which immediately yield from
definition (4) the ghost regime $\omega<-3/2$. This regime naturally is in
contradiction to the range $-3/2<\omega<-4/3$ derived in [13].

We wish to state the following (for the definition of real constants $C$,
$\omega$ and $\lambda$ referred to below, see next section):

(1) Visser and Hochberg [2] detailed the geometry of vacuum Brans Class I
wormhole in JF for different ranges of $\omega$ ($\omega<-2$ \textit{as well
as} $\omega>-3/2$). We want to emphasize that the new finite regime that we
derived and studied ($-3/2<\omega<-4/3$) is a result of our JF wormhole
condition $(C+1)^{2}>\lambda^{2}$, which is \textit{weaker} than conditions on
$B_{\text{VH}}=-\frac{C+1}{\lambda}$. To be more precise, Visser and Hochberg
[2] used only $\lambda>0$, whereas we permitted $\lambda$ to have both signs
allowing us to carve out a narrow, but new, regime for $\omega$ at the expense
of sacrificing the general relativity limit at $\omega\rightarrow\pm\infty$.
(It is not even any sacrifice, as is known today, see last few references
cited in [13] and Sec.IV below).

(2) What is the basis for our derived range $-3/2<\omega<-4/3$? Note that from
most general considerations, the weak field value of $C$ is $C=-\frac
{1}{\omega+2}$, which yields $\lambda=\pm\sqrt{\frac{2\omega+3}{2\omega+4}}$.
The \textit{reality} of $\lambda$ suggests that $-3/2<\omega$. The other
limit, $\omega<-4/3$, is obtained by imposing on the solution the fundamental
wormhole constraint $(C+1)^{2}>\lambda^{2}$ and requiring further that the
throat be real. Nowhere was it necessary to appeal to the transformations
(3),(4). The parameters we used in [13] were all from within the JF theory,
where $\omega$ is understood with its full physical meaning. The singularity
at $r=B$ was only too evident, but we relied on the fact that the traveler had
no access to it.

(3) The no-wormhole theorem is based on mapping between JF and EF via
conformal transformations whereas we know that conformally related geometries
are not the same geometries; for instance, curved cosmological metrics can be
conformally flat. There are other reasons too, some of which are well known:
(i) There is ostensibly no $\omega$ in the minimally coupled EF action, a
theory that can stand independently by itself without any umbilical relation
to JF. (ii) The minimally coupled theory is \textit{no} Machian Brans-Dicke
scalar-tensor theory, but already the non-Machian Einstein's general
relativity with a material source term. (iii) The vacuum JF theory is
\textit{conformally invariant} while EF theory is not (See Sec.IV). (iv)
Contrary to prevailing belief, it is impossible to obtain Einstein's theory in
the $\left\vert \omega\right\vert \rightarrow\infty$ limit of the vacuum
Brans-Dicke theory [17]. (v) The kinetic term (giving ghost or no-ghost) is
identifiable as a source only in the EF action (5), while it is unidentifiable
in the JF action (1) and finally (vi) The two frames can also be
observationally distinguishable [18].

Despite the above, some authors consider the Einstein minimally coupled theory
(5) exactly the same as Brans-Dicke theory (1), only rephrased in the Einstein
frame. We disagree with this point of view. Some authors regard the EF as a
convenient mathematical tool to draw conclusions about JF. While this
procedure could possibly be useful for stability analysis, it could be
misleading as well. For instance, the conditions $\alpha>0$, $\omega>-3/2$,
$\phi$ real lead to a positive sign kinetic term (i.e., no ghost, no energy
condition violation) in the EF action (5), but they would hide the fact that
in JF, nonetheless, the energy conditions are violated, real throats do exist
for $\omega>-3/2$ in the Class I solution. But the hope for a true wormhole is
dashed by the occurrence of singularities both \ in JF and EF version of the
solution. The purpose of these arguments is to make a case for the principle
that one should look for true wormhole $\omega-$regime staying in any one
frame, and not make a detour into the other, when drawing physical conclusions
about solutions in that frame.

We next turn to criticism (B) that has in fact prodded us to look for regular
wormholes in the JF, if any. As mentioned before, the appearance of naked
singularity in Brans Class I spacetime is a genuine problem \ because it does
not make the solution in its usual form look like a wormhole worth its name.
We resolve the problem by certain explicit operations on the solution, as
shown below.

\textbf{III. Brans Class I solution}

The general solution of the field equations (2), in isotropic coordinates
($t,r,\theta,\psi$), is given by%
\begin{equation}
d\tau^{2}=-e^{2\alpha(r)}dt^{2}+e^{2\beta(r)}[dr^{2}+r^{2}(d\theta^{2}%
+\sin^{2}\theta d\psi^{2})].
\end{equation}
Brans class I solution [19] in JF is given by%
\begin{equation}
e^{\alpha(r)}=e^{\alpha_{0}}\left[  \frac{1-B/r}{1+B/r}\right]  ^{\frac
{1}{\lambda}},
\end{equation}%
\begin{equation}
e^{\beta(r)}=e^{\beta_{0}}\left[  1+B/r\right]  ^{2}\left[  \frac
{1-B/r}{1+B/r}\right]  ^{\frac{\lambda-C-1}{\lambda}},
\end{equation}%
\begin{equation}
\varphi(r)=\varphi_{0}\left[  \frac{1-B/r}{1+B/r}\right]  ^{\frac{C}{\lambda}%
},
\end{equation}%
\begin{equation}
\lambda^{2}\equiv(C+1)^{2}-C\left(  1-\frac{\omega C}{2}\right)  >0,
\end{equation}
where $\lambda$, $\alpha_{0}$, $\beta_{0}$, $B$, $C$, and $\varphi_{0}$ are
real constants. The constants $\alpha_{0}$ and $\beta_{0}$ are determined by
asymptotic flatness at $r=\infty$ as $\alpha_{0}=$ $\beta_{0}=0$.

To see the naked singularity at $r=B$, it is enough to consider the invariant
Riemann curvature component in the freely falling orthonormal frame
($\widehat{e}_{0^{\prime},}\widehat{e}_{1^{\prime},}\widehat{e}_{2^{\prime}%
,}\widehat{e}_{3^{\prime}}$) (See [13] for details)
\begin{equation}
\mathbf{R}_{\widehat{1}^{\prime}\widehat{0}^{\prime}\widehat{1}^{\prime
}\widehat{0}^{\prime}}=\frac{4Br^{3}Z^{2}[\lambda(r^{2}+B^{2})-Br(C+2)]}%
{\lambda^{2}(r^{2}-B^{2})^{4}},
\end{equation}
where%
\begin{equation}
Z\equiv\left(  \frac{r-B}{r+B}\right)  ^{(C+1)/\lambda}.
\end{equation}
Clearly, $\mathbf{R}_{\widehat{1}^{\prime}\widehat{0}^{\prime}\widehat
{1}^{\prime}\widehat{0}^{\prime}}\rightarrow\infty$ as $r\rightarrow B$. All
curvature invariants also exhibit this behavior. To remove this singularity,
we do the following operations on the Class I solution [21]:%
\begin{equation}
r\rightarrow\frac{1}{r^{\prime}}\text{, }B\rightarrow\frac{i}{B^{\prime}%
}\text{, }\lambda\rightarrow-i\Lambda\text{, }\alpha_{0}\rightarrow\text{
}\epsilon_{0}\text{, }\beta_{0}\rightarrow\zeta_{0}+2\ln B^{\prime},
\end{equation}
where $B^{\prime}$, $\Lambda$ are real. Using the identity%
\begin{equation}
\tan^{-1}(x)=\frac{i}{2}\ln\left(  \frac{1-ix}{1+ix}\right)  ,
\end{equation}
we arrive at the metric functions and the scalar field as follows%
\begin{equation}
d\tau^{2}=-e^{2\alpha(r^{\prime})}dt^{2}+e^{2\beta(r^{\prime})}[dr^{\prime
2}+r^{\prime2}(d\theta^{2}+\sin^{2}\theta d\psi^{2})],
\end{equation}
where
\begin{equation}
\alpha(r^{\prime})=\epsilon_{0}+\frac{2}{\Lambda}\tan^{-1}\left(
\frac{r^{\prime}}{B^{\prime}}\right)
\end{equation}%
\begin{equation}
\beta(r^{\prime})=\zeta_{0}-\frac{2(C+1)}{\Lambda}\tan^{-1}\left(
\frac{r^{\prime}}{B^{\prime}}\right)  -\ln\left(  \frac{r^{\prime2}}%
{r^{\prime2}+B^{\prime2}}\right)
\end{equation}%
\begin{equation}
\varphi(r^{\prime})=\varphi_{0}\exp\left[  \frac{2C}{\Lambda}\tan^{-1}\left(
\frac{r^{\prime}}{B^{\prime}}\right)  \right]
\end{equation}%
\begin{equation}
\Lambda^{2}\equiv C\left(  1-\frac{\omega C}{2}\right)  -(C+1)^{2}>0.
\end{equation}
Asymptotic flatness requires that $\epsilon_{0}=-\frac{\pi}{\Lambda}$ and
$\zeta_{0}=\frac{\pi(C+1)}{\Lambda}$. This form of Class I solution has been
listed by Brans [19] as his Class II solution, but we see that the two classes
are \textit{not} independent $-$ one can be derived from the other by Wick
rotation. Likewise, we show that the singular Ellis I and the non-singular
Ellis III solutions in the minimally coupled theory are not independent
solutions (See Appendix for an outline). \textit{The main conclusion is that
the solution set (15)-(19) is the counterpart in JF of the well known Ellis
III wormhole in EF.}

The solution set (15)-(19) is regular everywhere including at $r^{\prime
}=B^{\prime}$ as can be verified by computing the curvature invariants. It has
two asymptotically flat regions\footnote{But not inversion invariant due to
asymmetry, like in the Ellis III wormhole.} with two asymmetric masses
$\frac{2B^{\prime}C}{\Lambda}$ and $-\frac{2B^{\prime}C}{\Lambda}$exp$\left[
-\frac{\pi}{\Lambda}\right]  $ on either side connected by a throat at%
\begin{equation}
r_{0}^{\prime\pm}=B^{\prime}\left[  \frac{C+1}{\Lambda}\pm\sqrt{1+\left(
\frac{C+1}{\Lambda}\right)  ^{2}}\right]  ,
\end{equation}
defined by the minimum of the area radius $r^{\prime}\exp[\beta(r^{\prime})]$.
The Riemann curvature component invariant under Lorentz boost is
\begin{equation}
\mathbf{R}_{\widehat{1}^{\prime}\widehat{0}^{\prime}\widehat{1}^{\prime
}\widehat{0}^{\prime}}=-\frac{4B^{\prime5}r^{\prime3}[\Lambda B^{\prime
2}+B^{\prime}r^{\prime}(C+2)-\Lambda r^{\prime2}]}{\Lambda^{2}(r^{\prime
2}+B^{\prime2})^{4}}\text{exp}\left[  \frac{8(C+1)}{\Lambda}\tan^{-1}\left(
\frac{r^{\prime}}{B^{\prime}}\right)  \right]  ,
\end{equation}
which is finite everywhere, and $\mathbf{R}_{\widehat{1}^{\prime}\widehat
{0}^{\prime}\widehat{1}^{\prime}\widehat{0}^{\prime}}\rightarrow0$ as
$r^{\prime}\rightarrow\pm\infty$. All the curvature invariants are also finite
and go to zero as $r^{\prime}\rightarrow\pm\infty$. These facts resolve the
maladies associated with the original form of Class I solution.

Using the weak field value $C=-\frac{1}{\omega+2}$, we get%
\begin{equation}
\Lambda=\pm\sqrt{-\frac{2\omega+3}{2\omega+4}}.
\end{equation}
To analyze the behavior of curvature $\mathbf{R}_{\widehat{1}^{\prime}%
\widehat{0}^{\prime}\widehat{1}^{\prime}\widehat{0}^{\prime}}$ or the radial
tidal force, we first implement that $\Lambda$ be real, which immediately
yields a new range $-2<\omega<-3/2$. Putting the values of $C$ and $\Lambda$,
together with any value of $\omega$ in the said range, we would obtain two
values for the throat $r_{0}^{\prime\pm}$, one positive and the other
negative, for real $B^{\prime}$. Next, we shall discard the negative value for
$r_{0}^{\prime}$ due to the fact that it would correspond to negative
circumferential radius $2\pi R_{0}$ for the throat defined in the generic
Morris-Thorne form\footnote{The radial coordinate $r^{\prime}$ is an abstract
coordinate chart covering the entire space, whereas the Morris-Thorne radius
$R$ is defined by physically measurable circumference but it does not cover
the entire space. The throat radius can be calculated either by the minimal
area radius involving $r^{\prime}$ or from the shape function involving $R$.
Both of course yield the same answer.}:
\begin{equation}
R_{0}=r_{0}^{\prime}\left[  1+\frac{B^{\prime2}}{r_{0}^{\prime2}}\right]
\exp\left[  \zeta_{0}-\frac{2(C+1)}{\Lambda}\tan^{-1}\left(  \frac
{r_{0}^{\prime}}{B^{\prime}}\right)  \right]  .
\end{equation}
Putting the value of $C$ and either value of $\Lambda$ in turn we can express
$\mathbf{R}_{\widehat{1}^{\prime}\widehat{0}^{\prime}\widehat{1}^{\prime
}\widehat{0}^{\prime}}=g\mathbf{(}\omega,r^{\prime},B^{\prime})$, where the
function $g$ results from the right hand side of (21). Finally, the behavior
of $g$ in the figures 1 and 2 exhibit the wormhole analogue of the naked black
hole. For positive $\Lambda$, curvature increases, while for negative
$\Lambda$, curvature depletes above the throat, as measured by a Lorentz
boosted observer. In either case, the hump and dip in the plots show that the
curvature function $g$ does not monotonically increase or decrease near the
throat, which resemble the phenomena occurring near the horizon in naked black holes.

\textbf{IV. Conformal invariance of vacuum Brans-Dicke theory}

We point out an important fact about JF vacuum Brans-Dicke theory. Under
conformal transformations [17,20,22]%
\begin{equation}
\widetilde{g}_{\mu\nu}=\varphi^{2\xi}g_{\mu\nu}%
\end{equation}
and a redefinition of the scalar%
\begin{equation}
\sigma=\varphi^{1-2\xi},
\end{equation}
the Brans-Dicke action (1) remains invariant%
\begin{equation}
S=\int d^{4}x(-\widetilde{g})^{1/2}\left[  \sigma\widetilde{\mathbf{R}%
}+\widetilde{\omega}\sigma^{-1}\widetilde{g}^{\mu\nu}\sigma_{,\mu}\sigma
_{,\nu}\right]  ,
\end{equation}
where $\xi$ is a real gauge parameter and \ \
\begin{equation}
\widetilde{\omega}=\frac{\omega-6\xi(\xi-1)}{(1-2\xi)^{2}}.
\end{equation}

The invariance means that the vacuum solutions are not unique. Given any
solution $(g_{\mu\nu},\varphi,\omega)$, it is possible to generate any other
solution $(\widetilde{g}_{\mu\nu},\sigma,\widetilde{\omega})$. The
transformations (24), (25), collectively denoted by $T_{\xi}$, form a
1-parameter Abelian group for $\xi\neq\frac{1}{2}$ and all the Brans-Dicke
manifolds $(g_{\mu\nu},\varphi,\omega)$ mapped by $T_{\xi}$ into
$(\widetilde{g}_{\mu\nu},\sigma,\widetilde{\omega})$ form an equivalence
class
%TCIMACRO{\tciFourier}%
%BeginExpansion
$\mathcal{F}$%
%EndExpansion
. The effect of $T_{\xi}$ is to move $\omega$ into another value
$\widetilde{\omega}$ depending on the choice of gauge $\xi$, except the
identity transformation $\omega=\widetilde{\omega}$ at $\xi=0,1$. \textit{The
}$\omega\rightarrow\infty$\textit{ limit can also be seen as a parameter
change that moves Brans-Dicke theory within the same class
%TCIMACRO{\tciFourier}%
%BeginExpansion
$\mathcal{F}$%
%EndExpansion
, and therefore it cannot reproduce Einstein's general relativity, which does
not belong to
%TCIMACRO{\tciFourier}%
%BeginExpansion
$\mathcal{F}$%
%EndExpansion
.} The conformal invariance is broken when ordinary matter is added to the
action except when the trace $\mathbf{T}^{(\text{matter})}=0$. When
$\xi\rightarrow\frac{1}{2}$, one has $\widetilde{\omega}\rightarrow\infty$,
the behavior of $\sigma$ becomes problematic\footnote{The prevailing belief
has been that the Brans-Dicke scalar field $\sigma$ possesses the asymptotic
behavior $\sigma=\sigma_{0}+O(\frac{1}{\widetilde{\omega}})$ so that one
recovers general relativity in the limit $\widetilde{\omega}\rightarrow\infty
$. This is not the case; it instead shows the asymptotic behaviour
$\sigma=\sigma_{0}+O\left(  \frac{1}{\sqrt{\widetilde{\omega}}}\right)  $. To
see it explicitly, take any starting finite value $\omega<-3/2$ and expand
$\xi\simeq\frac{1}{2}\left(  1\pm\frac{\sqrt{9+6\widetilde{\omega}%
+6\omega+4\omega\widetilde{\omega}}}{3+2\widetilde{\omega}}\right)  $. To
ensure the reality of $\xi$, we would require that $\widetilde{\omega}<-3/2$
as well. When $\xi\rightarrow1/2$, $\widetilde{\omega}\rightarrow-\infty$, we
have $\sigma\simeq1\mp\sqrt{\frac{3+2\omega}{2\widetilde{\omega}}}\ln\varphi$
so that $\sigma_{,\mu}\simeq\mp\sqrt{\frac{3+2\omega}{2\widetilde{\omega}}%
}\left(  \ln\varphi\right)  _{,\mu}$. Then the first term on the right hand
side of (2) does not vanish in the limit $\widetilde{\omega}\rightarrow
-\infty$. This is the source of trouble.
\par
.}. One could instead use the redefinition (4) but then the parameter $\omega$
disappears altogether in (5), and the $\omega\rightarrow\infty$ limit cannot
be considered because the theory is already GR, apart from a possible
violation of the equivalence principle due to the anomalous coupling of the
scalar to the energy--momentum tensor of ordinary matter, if $T_{\mu\nu}\neq
0$. It is evident from (27) that a fixed finite value of $\omega$ is not
sacrosanct $-$ it can be moved to any desired value $\widetilde{\omega}$
without getting out of the JF Brans-Dicke theory. The value $\omega
=\widetilde{\omega}$ $=-3/2$ is a fixed point of the transformation (27). Any
value of $\omega\lessgtr-3/2$ can be moved to any other value $\widetilde
{\omega}\,\lessgtr-3/2$ by choosing the gauge parameter $\xi$. However, it is
not possible to move any value of $\omega<-3/2$ across the divide
$\omega=\widetilde{\omega}$ $=-3/2$ to any other value $\widetilde{\omega
}\,>-3/2$ and vice versa because the gauge parameter $\xi$ becomes complex.

\textbf{V. Summary}

The present article has been induced by the two articles [14], [15] that deal
with the more general problem of $f(R)$ gravity and general scalar-tensor
theories. We have been concerned here only with the special case of $\omega=$
constant vacuum Brans-Dicke theory and limited our analysis to the critical
remarks against the range ($-3/2<\omega<-4/3$) in [15, pp.2,7]. We agree that
this range does not provide the ghost kinetic term via the mapping (4), hence
wormhole, \textit{in }EF and acknowledge the well known singularity problem in
the Class I solution in JF.

We have argued that the two frames, JF and EF, are physically distinct and
distinguishable; exoticity in one frame need not lead to exoticity in the
other. Because of this, it makes sense to search for regular wormholes working
only within any one of the frames. The old range ($-3/2<\omega<-4/3$) in the
Class I solution did yield some of the wormhole features except regularity
[13]. In fact, the solution represents two separate singular space-times, not
two regions of one \textit{connected }space-time required of a true wormhole.
Our principle has been, both in [13] and here, to work solely within the JF
without needing any mapping to EF. Accordingly, in the present article, we
looked for a regular Brans-Dicke wormhole with two asymptotically flat regions
of one connected spacetime and found it in (15)-(19) from Class I solution via
operations (13). All the features of this wormhole can be readily transferred
to EF, if necessary.

We have found that the operations (13) on the Brans Class I solution do remove
the singularity in it. The transformed solution can be recognized as Brans
Class II solution, which is the JF counterpart of the well known regular Ellis
III wormhole in EF. As we see, the Brans Classes of solutions I and II are not
independent, which seems to be a less known fact. Parallel to it, we show in
Appendix that the Ellis I and III solutions in the minimally coupled theory
are also not independent [23]. Our main conclusion is that the
\textit{transformed} Brans Class I solution (15)-(19) is a regular, twice
asymptotically flat wormhole with all other desirable properties in the range
$-2<\omega<-3/2$. This incidentally answers the remark in [15] about this interval.

The conformal invariance of the vacuum JF theory shows that the solutions in
it are not unique. Varying the real gauge parameter $\xi$, one can obtain any
value $\widetilde{\omega}$ from a given $\omega$ on either side of the divide
$\omega=-3/2$ but not\textit{ across} it since it is the fixed point of the
transformation (27). Thus our new range $-2<\omega<-3/2$ is based on the gauge
$\xi=0$, while the lower limit ($-2<\omega$) can always be shifted to any
value $\widetilde{\omega}$ for suitable $\xi\neq0$. We confirm that the
wormhole analogue of naked black holes exist also in the regular wormhole
spacetime, as evident from figures 1 and 2.

\textbf{Acknowledgments}

KKN\ sincerely thanks Professor Alexei A. Starobinsky for critical
correspondences motivating the present investigation. We also thank Professors
Alexander N. Petrov, Alexander I. Filippov, Olga and Nicolai Mikolaychuk for
many useful discussions. We are indebted to Guzel N. Kutdusova for helping us
with the language syntax.

\begin{center}
\textbf{Appendix }
\end{center}

The following solution follows directly from Brans Class I solution under
conformal transformation [4]:%

\begin{align}
d\tau_{E}^{2} &  =\widetilde{g}_{\mu\nu}dx^{\mu}dx^{\nu}=-\left(
\frac{1-\frac{m}{2r}}{1+\frac{m}{2r}}\right)  ^{2\beta}dt^{2}+\left(
1-\frac{m}{2r}\right)  ^{2(1-\beta)}\left(  1+\frac{m}{2r}\right)
^{2(1+\beta)}\times\nonumber\\
&  [dr^{2}+r^{2}d\theta^{2}+r^{2}\sin^{2}\theta d\psi^{2}]\tag{A1}%
\end{align}

\begin{equation}
\phi(r)=\sqrt{\frac{2(1-\beta^{2})}{\alpha}}\ln\left[  \frac{1-\frac{m}{2r}%
}{1+\frac{m}{2r}}\right]  ,\tag{A2}%
\end{equation}
of where $\alpha$, $m$ and $\beta$ are arbitrary positive constants. This is
known as Buchdahl solution [24] of Einstein minimally coupled theory (EF),
rediscovered later as Ellis I solution. The metric is invariant in form under
inversion (for integer $\beta$) of the radial coordinate $r\rightarrow
\frac{m^{2}}{4r}$ and we have two asymptotically flat regions (at $r=0$ and
$r=\infty$), the minimum area radius (throat) occurring at $r_{0}=\frac{m}%
{2}\left[  \beta+\sqrt{\beta^{2}-1}\right]  $. The real throat is guaranteed
by the condition $\alpha=-1$, $\beta^{2}>1$, leading to negative kinetic term.
Once again, for $\beta=1$, it reduces to the Schwarzschild black hole solution
in isotropic coordinates and for $\beta^{2}>1$, it represents a naked
singularity at $r=m/2$.

In the above solution, ghost matter exists as it can be easily seen that NEC
and WEC are violated. However, despite having two flat asymptotic regions, the
solution is spoilt because of the occurrence of singularity, though a traveler
would never meet it.

Using the coordinate transformation $l=r+\frac{m^{2}}{4r}$, the solution (A1)
and (A2) can be expressed as%
\begin{align}
ds^{2} &  =-f_{0}(l)dt^{2}+\frac{1}{f_{0}(l)}\left[  dl^{2}+(l^{2}%
-m^{2})\left(  d\theta^{2}+\sin^{2}\theta d\psi^{2}\right)  \right]
,\tag{A3}\\
f_{0}(l) &  =\left(  \frac{l-m}{l+m}\right)  ^{\beta},\tag{A4}\\
\phi_{0}(l) &  =\sqrt{\frac{\beta^{2}-1}{2}}\ln\left[  \frac{l-m}{l+m}\right]
.\tag{A5}%
\end{align}
In the solution set ($f_{0}$,$\phi_{0}$), we choose
\begin{equation}
m\rightarrow-im,\beta\rightarrow i\beta\tag{A6}%
\end{equation}
so that the throat $l_{0}=m\beta$ remains invariant in sign and magnitude.
Then the metric resulting from (A3) is
\begin{equation}
ds^{2}=-f_{0}^{\prime}(l)dt^{2}+\frac{1}{f_{0}^{\prime}(l)}\left[
dl^{2}+(l^{2}+m^{2})\left(  d\theta^{2}+\sin^{2}\theta d\psi^{2}\right)
\right]  \tag{A7}%
\end{equation}%
\begin{equation}
f_{0}^{\prime}(l)=\exp\left[  -2\beta\cot^{-1}\left(  \frac{l}{m}\right)
\right]  \tag{A8}%
\end{equation}%
\begin{equation}
\phi_{0}^{\prime}(l)=\left[  \sqrt{2}\sqrt{1+\beta^{2}}\right]  \cot
^{-1}\left(  \frac{l}{m}\right)  .\tag{A9}%
\end{equation}
Using the relation
\begin{align}
\cot^{-1}(x)+\tan^{-1}(x) &  =+\frac{\pi}{2};x>0\tag{A10}\\
&  =-\frac{\pi}{2};x<0\tag{A11}%
\end{align}
we get from (A8), (A9)%
\begin{equation}
f_{0\pm}(l)=\exp\left[  -2\beta\left\{  \pm\frac{\pi}{2}-\tan^{-1}\left(
\frac{l}{m}\right)  \right\}  \right]  \tag{A12}%
\end{equation}%
\begin{equation}
\phi_{0\pm}(l)=\left[  \sqrt{2}\sqrt{1+\beta^{2}}\right]  \left[  \pm\frac
{\pi}{2}-\tan^{-1}\left(  \frac{l}{m}\right)  \right]  \tag{A13}%
\end{equation}
We might study the solutions (A8) and (A9) \textit{per se}, while allowing for
a discontinuity at the origin $l=0$. Alternatively, we might treat
\textit{each} of the $\pm$ set in Eqs.(A12), (A13) as independently derived
exact solution valid in the unrestricted range of $l$ with no discontinuous
jump at $l=0$. The two alternatives do not appear quite the same. In fact,
each of the individual branch represents a geodesically complete,
asymptotically flat asymmetric wormhole having different masses, one positive
($m\beta$) and the other negative ($-m\beta e^{-\beta\pi}$), on two sides
respectively. The commonly known Ellis III solution is the $+ve$ branch which
is continuous over the entire interval $l\in(-\infty,+\infty)$. The $-ve$
branch also possesses exactly the same properties. What we have shown here is
that the Ellis I\ and III solutions are \textit{not} independent solutions of
the Einstein minimally coupled scalar field theory. The Brans Class II
wormhole is the JF counterpart of, but not the same as, Ellis III
wormhole\footnote{This is referred to also as Ellis-Bronnikov solution in
[23]. The solutions of Einstein minimally coupled theory was first found by
Fisher [25] in 1948 in a certain form, rediscovered by several authors
afterwards differing only in coordinate choices. The notable authors, among
others, include Buchdahl [24], Ellis [26] and Bronnikov [27]. We have shown
that two classes of solutions can be derived from one another and are not
really independent solutions.
\par
{}} since conformal mapping and Wick rotations do not commute.

\textbf{Figure captions}

Fig.1. We take $B^{\prime}=1$, the positive sign before $\Lambda$ and a value
in the new range $-2<\omega<-3/2$, say, $\omega=-1.7$. The positive value of
the throat radius is $r_{0}^{\prime+}=0.17$, $C=-3.33$ and $\Lambda=0.81$. The
maximum value of $\mathbf{R}_{\widehat{1}^{\prime}\widehat{0}^{\prime}%
\widehat{1}^{\prime}\widehat{0}^{\prime}}$ occurs at $r^{\prime}=0.54$, which
lies above the throat $r_{0}^{\prime+}$. The plot shows curvature enhancement
above the throat.

Fig.2. We take $B^{\prime}=1$, the negative sign before $\Lambda$ and a value
in the new range $-2<\omega<-3/2$, say, $\omega=-1.7$. The positive value of
the throat radius is $r_{0}^{\prime+}=5.88$, $C=-3.33$ and $\Lambda=-0.81$.
The minimum value of $\mathbf{R}_{\widehat{1}^{\prime}\widehat{0}^{\prime
}\widehat{1}^{\prime}\widehat{0}^{\prime}}$ occurs at $r^{\prime}=8.60$, which
lies above the throat $r_{0}^{\prime+}$. The plot shows curvature depletion
above the throat.

\textbf{References}

[1] A. G. Agnese and M. La Camera, Phys. Rev. D \textbf{51}, 2011 (1995).

[2] M. Visser and D. Hochberg, \textit{Proc. Haifa Workshop on the Internal
Structure of Black Holes and Space Time Singularities }(Jerusalem, Israel,
June, 1997) [arXiv:gr-qc/970001], p.20.

[3] K.K. Nandi, A. Islam and J. Evans, Phys. Rev. D \textbf{55}, 2497 (1997).

[4] K. K. Nandi, B. Bhattacharjee, S. M. K. Alam, J. Evans, Phys. Rev. D
\textbf{57}, 823 (1998).

[5] Luis J. Garay and Juan Garcia-Bellido, Nucl. Phys. B \textbf{400}, 416 (1993).

[6] Luis A. Anchordoqui, Santiago Perez Bergliaffa and Diego F. Torres, Phys.
Rev. D \textbf{55}, 5226 (1997).

[8] Arunava Bhadra, Ion Simaciu, Kamal Kanti Nandi and Yuan-Zhong Zhang, Phys.
Rev. D \textbf{71, }128501 (2005).

[9] Arunava Bhadra and Kabita Sarkar, Mod. Phys. Lett. A \textbf{20, }1831(2005).

[10] Ernesto F. Eiroa, Martin G. Richarte and Claudio Simeone, Phys. Lett. A
\textbf{373}, 1 (2008); ibid. A \textbf{373}E, 2399 (2009).

[11] S.M.Kozyrev and S. V. Sushkov [arXiv:gr-qc/0812.5010]

[12] Francisco S. N. Lobo and Miguel A. Oliveira, Phys. Rev.D \textbf{81},
067501 (2010).

[13] Amrita Bhattacharya, Ilnur Nigmatzyanov, Ramil Izmailov and Kamal K.
Nandi, Class.Quant.Grav. \textbf{26}, 235017 (2009).

[14] Kirill A. Bronnikov and Alexei A. Starobinsky, JETP Lett. \textbf{85},1 (2007).

[15] K.A. Bronnikov, M.V. Skvortsova and A.A. Starobinsky,
[arXiv:gr-qc/1005.3262v1](To appear in Grav. \& Cosmol.).

[16] M. Visser, \textit{Lorentzian Wormholes---From Einstein To Hawking} (AIP,
New York, 1995).

[17] V. Faraoni, Phys. Lett. A \textbf{245}, 26 (1998).

[18] A. Bhadra, K. Sarkar, D. P. Datta and K. K. Nandi, Mod. Phys. Lett. A
\textbf{22, }367 (2007).

[19] C. H. Brans, Phys. Rev. \textbf{125}, 2194 (1962).

[20] A. Bhadra and K.K. Nandi, Phys. Rev. D \textbf{64}, 087501(2001).

[21] Arunava Bhadra and Kabita Sarkar, Mod. Phys. Lett. A \textbf{20}, 1831 (2005).

[22] Y.M. Cho, Phys. Rev. Lett. \textbf{68}, 3133 (1992).

[23] Kamal K. Nandi, Ilnur Nigmatzyanov, Ramil Izmailov and Nail G. Migranov,
Class.Quant.Grav. \textbf{25},165020 (2008).

[24] H.A. Buchdahl, Phys. Rev. \textbf{115}, 1325 (1959).

[25] I.Z. Fisher, Zh. Eksp. Teor. Fiz. \textbf{18}, 636 (1948) [gr-qc/991108].

[26] H.G. Ellis, J. Math. Phys. \textbf{14}, 104 (1973); ibid. \textbf{15},
520E (1974).

[27] K.A. Bronnikov, Acta Phys. Polon. B \textbf{4}, 251 (1973).

\begin{center}
---------------------------------------------------------------
\end{center}

\bigskip

\end{document}